# Analysis of Particle Transport in a Magnetophoretic Microsystem

E. P. Furlani

*Institute for Lasers, Photonics and Biophotonics,*
*University at Buffalo (SUNY), Buffalo, NY, 14260*

[*]E. P. Furlani is the corresponding author: email efurlani@buffalo.edu,

# Abstract

An analytical analysis is presented of the transport and capture of magnetic micro/nano-particles in a magnetophoretic microsystem that consists of an array of integrated soft-magnetic elements embedded beneath a microfluidic channel. The elements, which are polarized by a bias field, produce a nonuniform field distribution that gives rise to a force on magnetic particles within the microchannel. The equations governing particle motion are derived using analytical expressions for the dominant magnetic and fluidic forces. The magnetic force is obtained using an analytical expression for the field distribution in the microchannel combined with a linear magnetization model for the magnetic response of particles. The theory takes into account particle size and material properties, the bias field, the dimensions of the microchannel, the fluid properties, and the flow velocity. The equations of motion are solved to study particle transport and capture. The analysis indicates that the particles exhibit an oscillatory motion as they traverse the microsystem, and that a high capture efficiency can be obtained in practice.

# I. INTRODUCTION

Research for the development of magnetophoretic microsystems is substantial and growing rapidly due primarily to applications in microbiology and biomedicine. In applications such as bioseparation, immuno-assays, and drug delivery, magnetic particles are used to label and manipulate biomaterials (cells, enzymes, proteins, DNA), and to transport therapeutic drugs.[1] The rapid progress in these areas is due to several factors including advances in microfluidic technology, and the development of functionalized magnetic particles that can selectively label, isolate, and transport a target biomaterial.[2-4]

Magnetophoretic microsystems are well suited for bioapplications because they enable (i) fast reaction times, (ii) the analysis and monitoring of small samples (potentially picoliters), and (iii) the integration of "micro total analysis systems" ($\mu$TAS).[5,6] They have additional advantages in that the relatively low permeability of an aqueous medium enables efficient coupling between an applied field and the magnetically labeled material. Moreover, the relatively low intrinsic magnetic susceptibility of biomaterials provides substantial contrast between labeled and unlabeled material, which enables a high degree of detection and selectivity.

Several magnetophoretic microsystems have been developed for applications ranging from bioseparation to cell sorting.[7-12] In most of these systems the magnetic force is provided by integrated electromagnets. However, Han and Frazier have recently demonstrated a passive microsystem for sorting blood cells in which the field and force are provided by an integrated soft-magnetic element and an external bias field.[12] This system is designed to process biofluids containing both paramagnetic red blood cells and diamagnetic white blood cells. Specifically, when these two types of cells flow through the microsystem they experience a force of opposite polarity, that causes them to migrate towards separate outlet channels, thereby effecting cell sorting. The magnetic susceptibility of these cells is low, and therefore a large force is needed to implement separation. It is difficult to achieve a viable force for this application using electromagnets, as this would require a prohibitive level of current. In this regard, integrated soft-magnetic elements have advantages over electromagnets in that they (i) provide substantially more field and force for a given volume of material, (ii) consume no power, and (iii) generate no heat. As such, they hold great potential for a variety of existing and new magnetophoretic applications. However, despite their advantages, relatively few authors have considered particle transport in micosystems that utilize these elements. Instead, most of the theoretical work has been directed towards systems that use electromagnets.[8-10]

Particle transport in magnetophoretic micosystems has been studied using various theoretical approaches, depending on the complexity of the system. Typically, finite element analysis (FEA) is used to determine the magnetic field and force, and the particle motion is obtained using Newtonian analysis. However, numerical methods such as FEA are not well suited for the kind of parametric analysis that is needed for optimization, and are impractical for large-scale simulations. An analytical modeling approach is preferable for the development of new systems because it (i) provides insight into the fundamental behavior of the microsystem, (ii) enables the analysis of large-scale systems, (iii) provides exact solution values at each point in the solution space (as compared to mesh dependent interpolated or averaged values), and (iv) is well suited for parametric analysis.

In this article we use analytical analysis to study the transport and capture of magnetic micro/nano-particles in a passive magnetophoretic microsystem. The microsystem consists of an array of integrated soft-magnetic elements embedded in a nonmagnetic substrate beneath a microfluidic channel (Figs 1 and 2). A bias field is used to polarize the elements, which produce a field gradient that imparts a force to magnetic particles within the microchannel. We derive the equations of motion for the particles using analytical expressions for the dominant magnetic and fluidic forces. The magnetic force is obtained using an analytical expression for the field distribution in the microchannel, combined with a linear magnetization model with saturation for the magnetic response of particles. The theory takes into account key variables including the size and properties of the particles, the bias field, the dimensions of the microchannel, the fluid properties, and the flow velocity. We use these equations to study particle transport and capture in a practical microsystem. Our analysis demonstrates that the magnetic force within the microsystem is more than sufficient to enable efficient particle capture, which makes it a viable candidate for a variety of bioapplications. Moreover, our analysis further indicates that the particles exhibit an interesting oscillatory motion as they traverse the microsystem. This behavior could be exploited for a variety of applications, including the characterization of fluids at the micro or nanoscale.

## II. THEORY

Particle transport in a magnetophoretic system is governed by various forces including (a) the magnetic force due to all field sources, (b) viscous drag, (c) inertia, (d) gravity, (e) buoyancy, (f) thermal kinetics, (g) particle/fluid interactions (perturbations to the flow field), and (h) interparticle effects including; (i) magnetic dipole interactions, (ii) electric double-layer interactions, and (iii) van der Walls force. For most magnetophoretic applications involving submicron particles, the magnetic and viscous forces are dominant, and one can ignore all other effects. However, it is instructive to estimate the magnitude of the other forces for our application. To this end, we compute the gravitational and buoyant forces, $F_g = \rho \frac{4}{3}\pi R_p^3 g$ and $F_b = \rho_f \frac{4}{3}\pi R_p^3 g$, on a one micron ($R_p = 0.5\,\mu m$) iron-oxide (Fe$_3$O$_4$) particle in water ($\rho = 5000\,kg/m^3$, $\rho_f = 1000$ kg/m$^3$, and g = 9.8 m/s). We obtain $F_g$ = 2.618 × 10$^{-2}$ pN, and $F_b$ = 0.523 × 10$^{-2}$ pN, which are more than an order of magnitude smaller than the magnetic force (several pico-Newtons), as shown below. Thus, these forces can be neglected in our analysis. Similarly, the inertial force $m_p a$ is also a second order term and could be neglected, but we choose to include it in our analysis as noted below. As for the other forces, we assume that we are dealing with dilute particle suspensions (particle volume concentration c << 1), in which case, interparticle effects and particle/fluid interactions can also be neglected.[13] We further restrict the analysis to particles that are sufficiently large so that thermal kinetics (Brownian motion) can be ignored. For example, for Fe$_3$O$_4$ particles in water, Brownian motion is negligible when the particle diameter is greater than 40 nm.[14]

In this paper, we consider the transport of magnetic micro/nano-particles in slow flow regimes where the magnetic and viscous drag forces dominate, and where the motion of the particle can be predicted using Newton's second law,

$$m_p \frac{d\mathbf{v}_p}{dt} = \mathbf{F}_m + \mathbf{F}_f. \qquad (1)$$

Here, $m_p$ and $\mathbf{v}_p$ are the mass and velocity of the particle, and $\mathbf{F}_m$ and $\mathbf{F}_f$ are the magnetic and fluidic forces respectively. As noted above, the inertial term $m_p \frac{d\mathbf{v}_p}{dt}$ is small, and could be neglected, but we choose to include it to obtain more accurate particle trajectories during periods of high acceleration. However, an adequate description of the physics could be obtained using a simple force balance $\mathbf{F}_m = -\mathbf{F}_f$.

The magnetic force is given by

$$\mathbf{F}_m = \mu_0 V_p (\mathbf{M}_p \bullet \nabla) \mathbf{H}_a, \qquad (2)$$

where $\mu_0 = 4\pi \times 10^{-7}$ H/m is the permeability of free space, $\mathbf{V}_p$ and $\mathbf{M}_p$ are the volume and magnetization of the particle, and $\mathbf{H}_a$ is the applied magnetic field intensity at the center of the particle. The fluidic force is obtained using Stokes' law for the drag on a sphere,

$$\mathbf{F}_f = -6\pi\eta R_p (\mathbf{v}_p - \mathbf{v}_f), \tag{3}$$

where $\eta$ and $\mathbf{v}_f$ are the viscosity and the velocity of the fluid, respectively.

As noted above, Newton's equation (1) does not take into account Brownian motion, which can influence particle capture when the particle diameter $D_p$ is sufficiently small. Gerber et al.[14] have developed the following criterion to estimate this diameter,

$$|\mathbf{F}| D_p \leq kT, \tag{4}$$

where $|\mathbf{F}|$ is the magnitude of the total force acting on the particle, k is Boltzmann's constant, and T is the absolute temperature. For example, the threshold diameter for $Fe_3O_4$ particles in water is $D_p = kT/|\mathbf{F}| = 40$ nm. For particles smaller than this, one needs to solve an advection-diffusion equation for the particle number density $n(\mathbf{r},t)$, rather than Newton's equation for the trajectory of a single particle. The behavior of $n(\mathbf{r},t)$ is governed by the following equation,[14-16]

$$\frac{\partial n(\mathbf{r},t)}{\partial t} + \nabla \cdot \mathbf{J} = 0, \tag{5}$$

where $\mathbf{J} = \mathbf{J}_D + \mathbf{J}_F$ is the total flux of particles, which includes a contribution $\mathbf{J}_D = -D\nabla n(\mathbf{r},t)$ due to diffusion, and a contribution $\mathbf{J}_F = \mathbf{v} n(\mathbf{r},t)$ due to the action of all external forces. Equation (5) is often written in terms of the particle volume concentration $c(\mathbf{r},t)$, which is related to the number density, $c(\mathbf{r},t) = 4\pi R_p^3 n(\mathbf{r},t)/3$. The diffusion coefficient $D$ is given by the Nernst-Einstein relation $D = \mu kT$, where $\mu = 1/(6\pi\eta R_p)$ is the mobility of a particle of radius $R_p$ in a fluid with viscosity $\eta$ (Stokes' approximation).

The drift velocity also depends of the mobility $\mathbf{v}(\mathbf{r}) = \mu \mathbf{F}(\mathbf{r})$, where $\mathbf{F}(\mathbf{r}) = \mathbf{F}_m(\mathbf{r}) + \mathbf{F}_f(\mathbf{r}) + \cdots$, is the total force on the particle. Thus, Eq. (5) can be rewritten as

$$\frac{\partial n(\mathbf{r},t)}{\partial t} = \frac{kT}{(6\pi\eta R_p)}\nabla^2 n(\mathbf{r},t) - \frac{1}{(6\pi\eta R_p)}\nabla\cdot(\mathbf{F}(\mathbf{r})n(\mathbf{r},t)). \tag{6}$$

In order to solve either Newton's equation (1) or the advection-diffusion equation (6), one needs an expression for the forces $\mathbf{F}_m$ and $\mathbf{F}_f$, which we obtain below.

To determine the magnetic force $\mathbf{F}_m$ in Eq. (2), we use a linear magnetization model with saturation to predict the magnetization $\mathbf{M}_p$ of the particle, in which $\mathbf{M}_p$ is a linear function of the field up to a saturation value $M_{sp}$. Specifically, below saturation,

$$\mathbf{M}_p = \chi_p \mathbf{H}_{in}, \tag{7}$$

where $\mu_p$ and $\chi_p = \mu_p/\mu_0 - 1$ are permeability and susceptibility of the particle, and $\mathbf{H}_{in} = \mathbf{H}_a - \mathbf{H}_{demag}$, where $\mathbf{H}_{demag} = \mathbf{M}_p/3$ is the self-demagnetization field in the particle.[17] If the particle is suspended in a magnetically linear fluid of permeability $\mu_f$, the force on it is[18]

$$\mathbf{F}_m = \mu_f V_p \frac{3(\chi_p - \chi_f)}{\left[(\chi_p - \chi_f) + 3(\chi_f + 1)\right]}(\mathbf{H}_a \cdot \nabla)\mathbf{H}_a. \tag{8}$$

When $|\chi_f| \ll 1$ ($\mu_f \approx \mu_0$), (8) reduces to

$$\mathbf{F}_m = \mu_0 V_p \frac{3(\chi_p - \chi_f)}{(\chi_p - \chi_f) + 3}(\mathbf{H}_a \cdot \nabla)\mathbf{H}_a. \tag{9}$$

and

$$\mathbf{H}_{in} = \frac{3}{(\chi_p - \chi_f) + 3}\mathbf{H}_a, \tag{10}$$

and

$$\mathbf{M}_p = \frac{3(\chi_p - \chi_f)}{(\chi_p - \chi_f) + 3} \mathbf{H}_a. \quad (11)$$

The magnetization can be written as,

$$\mathbf{M}_p = f(\mathbf{H}_a)\mathbf{H}_a, \quad (12)$$

where

$$f(\mathbf{H}_a) = \begin{cases} \dfrac{3(\chi_p - \chi_f)}{(\chi_p - \chi_f) + 3} & \mathbf{H}_a < \left(\dfrac{(\chi_p - \chi_f) + 3}{3(\chi_p - \chi_f)}\right) \mathbf{M}_{sp} \\ \\ \mathbf{M}_{sp}/\mathbf{H}_a & \mathbf{H}_a \geq \left(\dfrac{(\chi_p - \chi_f) + 3}{3(\chi_p - \chi_f)}\right) \mathbf{M}_{sp} \end{cases}, \quad (13)$$

and $H_a = |\mathbf{H}_a|$.

The magnetic force can be decomposed into components,

$$\mathbf{F}_m(x,y) = F_{mx}(x,y)\hat{\mathbf{x}} + F_{my}(x,y)\hat{\mathbf{y}}, \quad (14)$$

where

$$F_{mx}(x,y) = \mu_0 V_p f(H_a)\left[H_{ax}(x,y)\frac{\partial H_{ax}(x,y)}{\partial x} + H_{ay}(x,y)\frac{\partial H_{ax}(x,y)}{\partial y}\right], \quad (15)$$

and

$$F_{my}(x,y) = \mu_0 V_p f(H_a)\left[H_{ax}(x,y)\frac{\partial H_{ay}(x,y)}{\partial x} + H_{ay}(x,y)\frac{\partial H_{ay}(x,y)}{\partial y}\right], \quad (16)$$

where

$$\mathbf{H}_a = H_{ax}(x,y)\hat{\mathbf{x}} + H_{ay}(x,y)\hat{\mathbf{y}}. \quad (17)$$

To evaluate $\mathbf{F}_m(x,y)$ we need an expression for the field distribution $\mathbf{H}_a(x,y)$. This is a superposition of two distinct fields, the bias field $\mathbf{H}_{bias}$, and the field $\mathbf{H}_e$ due to the array of magnetized soft-magnetic elements,

$$\begin{aligned} \mathbf{H}_a &= \mathbf{H}_{bias} + \mathbf{H}_e \\ &= \left(H_{bias,x} + H_{e,x}\right)\hat{\mathbf{x}} + \left(H_{bias,y} + H_{e,y}\right)\hat{\mathbf{y}}. \end{aligned} \tag{18}$$

However, $\mathbf{H}_{bias}$ and $\mathbf{H}_e$ are not both independent. Specifically, $\mathbf{H}_e$ depends on $\mathbf{H}_{bias}$ as it is the bias field that magnetizes the soft-magnetic elements. Therefore, $\mathbf{H}_{bias}$ induces $\mathbf{H}_e$.

**A. The Bias Field**

As noted above, a bias field is needed to magnetize the soft-magnetic elements in the microsystem so that they produce a local field gradient in the microchannel. Without this gradient, there is no magnetic force on the particles. The bias field also contributes to the magnetization of the particles. A strong bias field can be obtained by positioning rare-earth permanent magnets above and below the microsystem as shown in Fig. 1a. Permanent magnets have advantages over conventional electromagnets in that they (i) produce a substantially higher bias field for a given volume of material, (ii) consume no energy, and (iii) generate no heat.

The bias field can be predicted and optimized using analytical analysis. Consider the magnet shown in Fig. 3a. The bias field is the z-component of the magnetic field, which is given by[17]

$$\mathbf{B}_{z,bias}(x,y,z) = \frac{\mu_0 M_{bs}}{4\pi} \sum_{k=1}^{2}\sum_{n=1}^{2}\sum_{m=1}^{2} (-1)^{k+n+m} \tan^{-1}\left[\frac{(x-y_n)(y-y_m)}{(z-z_k)[(x-x_n)^2+(y-y_m)^2+(z-z_k)^2]^{1/2}}\right], \tag{19}$$

where $M_{bs}$ is the saturation magnetization of the magnet, and $x_n$, $y_m$ and $z_k$ ($n = m = k = 1, 2$) are the coordinates of its corners. It is important to note that the x, y, and z coordinates in (19) are with respect to a coordinate system fixed to the magnet as shown in Fig 4a, and are different than the coordinates used for the field and force analysis below.

The dimensions of the bias magnet are chosen to be much greater than those of the microsystem so that the bias field will be uniform across the microchannel. Rare-earth magnets such as neodymium-iron-boron (NdFeB) are ideal for this application. Consider a pair of identical NdFeB magnets positioned as shown in Fig. 1a. The saturation magnetization for medium grade NdFeB is $M_{bs} = 8.0 \times 10^5 \, A/m$. We use this value in (19) to compute the field distribution in the middle of the gap. Assume that the magnets have a cross-sectional area $A_{mag}$ = 20 mm × 20 mm facing the gap, and a height $h_{mag}$ = 20 mm. Further assume that the gap g = 4 mm, and that the microsystem is centered in the gap, 2 mm from the surface of either magnet. The field distribution in the middle of the gap at z = g/2 is shown in Fig. 4b. We spatially integrate this to obtain an average value $B_{z,ave}$ in the gap,

$$B_{z,ave} = \frac{1}{A_{mag}} \iint_{A_{mag}} \mathbf{B}_{z,bias}(x, y, g/2) \, dx \, dy . \qquad (20)$$

In this case, $B_{z,ave} = 0.55 \, T$, which is more than sufficient to saturate the soft-magnetic elements in the microsystem, as this is several orders of magnitude greater than the coercivity of materials such a permalloy that are commonly used for these applications.[17] In fact, for many applications, it sufficient to use a single bias magnet, and Eqs. (19) and (20) can be used to optimize its dimensions.

In summary, our analysis indicates that a bias field strong enough to saturate the soft-magnetic elements can easily be obtained using rare-earth permanent magnets. Thus, in the flowing sections we treat these elements as individual field sources with a fixed magnetization

.

## B. Field and Force of a Single Magnetic Element

Analytical expressions for the field and force of a single rectangular magnetic element of width 2w and height 2h that is centered with respect to the origin in the x-y plane (Fig. 4a) have been previously derived and verified using FEA.[17,19] They are repeated here for convenience. Assume that the element is magnetized to saturation $M_{es}$ by a bias field $H_{bias}$ in the vertical direction (along the y-axis). The field components are

$$H_{ax}^{(0)}(x,y) = \frac{M_{es}}{4\pi}\left\{\ln\left[\frac{(x+w)^2+(y-h)^2}{(x+w)^2+(y+h)^2}\right] - \ln\left[\frac{(x-w)^2+(y-h)^2}{(x-w)^2+(y+h)^2}\right]\right\}, \quad (21)$$

and

$$H_{ay}^{(0)}(x,y) = \frac{M_{es}}{2\pi}\left\{\tan^{-1}\left[\frac{2h(x+w)}{(x+w)^2+y^2-h^2}\right] - \tan^{-1}\left[\frac{2h(x-w)}{(x-w)^2+y^2-h^2}\right]\right\}. \quad (22)$$

The force components are

$$F_{mx}^{(0)}(x,y) = \mu_0 V_p f(H_a)\frac{M_{es}}{2\pi}\left\{\frac{M_{es}}{4\pi}\left(\ln\left[\frac{(x+w)^2+(y-h)^2}{(x+w)^2+(y+h)^2}\right] - \ln\left[\frac{(x-w)^2+(y-h)^2}{(x-w)^2+(y+h)^2}\right]\right)\right.$$

$$\times\left[\frac{x+w}{(x+w)^2+(y-h)^2} - \frac{x+w}{(x+w)^2+(y+h)^2} - \frac{x-w}{(x-w)^2+(y-h)^2} + \frac{x-w}{(x-w)^2+(y+h)^2}\right]$$

$$+ \left(H_{bias,y} + \frac{M_{es}}{2\pi}\left[\tan^{-1}\left[\frac{2h(x+w)}{(x+w)^2+y^2-h^2}\right] - \tan^{-1}\left[\frac{2h(x-w)}{(x-w)^2+y^2-h^2}\right]\right]\right)$$

$$\left.\times\left[\frac{y-h}{(x+w)^2+(y-h)^2} - \frac{y-h}{(x-w)^2+(y-h)^2} - \frac{y+h}{(x+w)^2+(y+h)^2} + \frac{y+h}{(x-w)^2+(y+h)^2}\right]\right\}, \quad (23)$$

and

$$F_{my}^{(0)} = \mu_0 V_p f(H_a) \frac{M_{es}}{\pi} \left\{ \frac{M_{es}}{4\pi} \left( \ln\left[\frac{(x+w)^2+(y-h)^2}{(x+w)^2+(y+h)^2}\right] - \ln\left[\frac{(x-w)^2+(y-h)^2}{(x-w)^2+(y+h)^2}\right] \right) \right.$$

$$\times \left[ \frac{h\left[y^2-(x+w)^2-h^2\right]}{\left[(x+w)^2+y^2-h^2\right]^2+4h^2(x+w)^2} - \frac{h\left[y^2-(x-w)^2-h^2\right]}{\left[(x-w)^2+y^2-h^2\right]^2+4h^2(x-w)^2} \right] \quad (24)$$

$$+ \left( H_{bias,y} + \frac{M_{es}}{2\pi} \left[ \tan^{-1}\left[\frac{2h(x+w)}{(x+w)^2+y^2-h^2}\right] - \tan^{-1}\left[\frac{2h(x-w)}{(x-w)^2+y^2-h^2}\right] \right] \right)$$

$$\left. \times \left[ \frac{2hy(x-w)}{\left[(x-w)^2+y^2-h^2\right]^2+4h^2(x-w)^2} - \frac{2hy(x+w)}{\left[(x+w)^2+y^2-h^2\right]^2+4h^2(x+w)^2} \right] \right\}$$

In deriving (23) and (24) we have used the following expressions for the field gradients,

$$\frac{\partial H_{ex}^{(0)}(x,y)}{\partial x} = \frac{M_{es}}{2\pi} \left\{ \frac{x+w}{(x+w)^2+(y-h)^2} - \frac{x+w}{(x+w)^2+(y+h)^2} \right.$$

$$\left. - \frac{x-w}{(x-w)^2+(y-h)^2} + \frac{x-w}{(x-w)^2+(y+h)^2} \right\}, \quad (25)$$

$$\frac{\partial H_{ex}^{(0)}(x,y)}{\partial y} = \frac{M_{es}}{2\pi} \left\{ \frac{y-h}{(x+w)^2+(y-h)^2} - \frac{y-h}{(x-w)^2+(y-h)^2} \right.$$

$$\left. - \frac{y+h}{(x+w)^2+(y+h)^2} + \frac{y+h}{(x-w)^2+(y+h)^2} \right\}, \quad (26)$$

and

$$\frac{\partial H_{ey}^{(0)}(x,y)}{\partial x} = \frac{M_{es}}{\pi} \left\{ \frac{h\left[y^2-(x+w)^2-h^2\right]}{\left[(x+w)^2+y^2-h^2\right]^2+4h^2(x+w)^2} \right.$$

$$\left. - \frac{h\left[y^2-(x-w)^2-h^2\right]}{\left[(x-w)^2+y^2-h^2\right]^2+4h^2(x-w)^2} \right\}, \quad (27)$$

$$\frac{\partial H_{ey}^{(0)}(x,y)}{\partial y} = \frac{2M_{es}}{\pi}\left\{\frac{hy(x-w)}{\left[(x-w)^2+y^2-h^2\right]^2+4h^2(x-w)^2}\right.$$

$$\left. - \frac{hy(x+w)}{\left[(x+w)^2+y^2-h^2\right]^2+4h^2(x+w)^2}\right\}. \quad (28)$$

We have also assumed that the bias field is constant and in the y-direction (no x-component),

$$\mathbf{H}_a = H_{e,x}\hat{x} + \left(H_{bias,y} + H_{e,y}\right)\hat{y}. \quad (29)$$

Therefore, $H_{bias}$ does not contribute to the gradient terms in (15) and (16). However, it is does contribute to the factor $f(H_a)$ in these expressions.

## C. Field and Force of a Magnetic Element Array

Analytical expressions for the field and force due to a linear array of soft-magnetic elements have also been previously derived, and are repeated here for convenience.[19] Let $N_e$ denote the number of elements in the array, and let the fist element be centered with respect to the origin in the x-y plane. All other elements are positioned along the x-axis as shown in Fig. 4b. We identify the elements using the index $n = (0,1,2,3,4, \ldots, N_e -1)$. The field components due to the first element ($n = 0$) are given by Eqs. (21), (22). The n'th element is centered at $x = s_n$ on the x-axis, and its field and force components can be written in terms of the 0'th components as follows:

$$H_{ex}^{(n)}(x,y) = H_{ex}^{(0)}(x-s_n,y) \quad H_{ey}^{(n)}(x,y) = H_{ey}^{(0)}(x-s_n,y) \quad (n=1,2,3,\ldots). \quad (30)$$

The total field of the array is obtained by summing the contributions from all the elements,

$$H_{ex}(x,y) = \sum_{n=0}^{N_e-1} H_{ex}^{(0)}(x-s_n,y), \quad (31)$$

$$H_{ey}(x,y) = \sum_{n=0}^{N_e-1} H_{ey}^{(0)}(x-s_n,y), \tag{32}$$

It follows from Eqs. (15) and (16) that the force components are given by.[19]

$$F_{mx}(x,y) = \mu_0 V_p f(H_a) \left[ \left( \sum_{n=0}^{N_e-1} H_{ex}^{(0)}(x-s_n,y) \right) \left( \sum_{n=0}^{N_e-1} \frac{\partial H_{ex}^{(0)}(x-s_n,y)}{\partial x} \right) \right.$$
$$\left. + \left( H_{bias,y} + \sum_{n=0}^{N_e-1} H_{ey}^{(0)}(x-s_n,y) \right) \left( \sum_{n=0}^{N_e-1} \frac{\partial H_{ex}^{(0)}(x-s_n,y)}{\partial y} \right) \right], \tag{33}$$

and

$$F_{my}(x,y) = \mu_0 V_p f(H_a) \left[ \left( \sum_{n=0}^{N_e-1} H_{ex}^{(0)}(x-s_n,y) \right) \left( \sum_{n=0}^{N_e-1} \frac{\partial H_{ey}^{(0)}(x-s_n,y)}{\partial x} \right) \right.$$
$$\left. + \left( H_{bias,y} + \sum_{n=0}^{N_e-1} H_{ey}^{(0)}(x-s_n,y) \right) \left( \sum_{n=0}^{N_e-1} \frac{\partial H_{ey}^{(0)}(x-s_n,y)}{\partial y} \right) \right]. \tag{34}$$

In Eqs. (33) and (34) we have assume that the bias field is constant and in the y-direction as in Eq. (29).

### D. Fluidic Force

To evaluate the fluidic force in Eq. (3) we need an expression for the fluid velocity $\mathbf{v}_f$ in the microchannel. Let L denote the length of the channel and $h_c$ and $w_c$ denote the half-height and half-width of its rectangular cross section (Fig. 2). The nature of the flow, laminar or turbulent, is estimated from the Reynolds number $Re = v_f D \rho / \eta$, where D is the characteristic length of the channel (the hydraulic diameter), and $\rho$ and $\eta$ are the density and viscosity of the fluid, respectively. In bioseparation applications V < 1 m/s, D ≈ 200 µm, $\rho \approx 1000$ kg/m$^3$, and $\eta \approx 0.001$ kg/m•s. Therefore, Re ≈ 200 which indicates laminar flow (i.e., Re < 2000). We assume fully developed laminar flow with the flow velocity parallel to the x-axis, and varying across its cross section,

$$\mathbf{v}_f = v_f(y', z')\,\hat{\mathbf{x}}. \tag{35}$$

It is convenient to use coordinates $y'$ and $z'$ centered with respect to the cross section of the channel, and it is understood that these differ from the coordinate system used for the magnetic analysis (Figs. 2 and 4). Here, $z'$ spans the width of the channel. The velocity profile for fully developed laminar flow is[20]

$$v_f(y',z') = \frac{16 h_c^2}{\eta \pi^3} \frac{\Delta P}{L} \sum_{n=0}^{\infty} \frac{(-1)^n}{(2n+1)^3} \left[1 - \frac{\cosh((2n+1)\pi z'/2h_c)}{\cosh((2n+1)\pi w_c/2h_c)}\right] \cos((2n+1)\pi y'/2h_c), \tag{36}$$

where $\Delta P$ is the change in pressure across the length L of the channel. The volume flow rate $Q$ through the channel is

$$Q = A\bar{v}_f, \tag{37}$$

where $A = 4h_c w_c$ is the cross-sectional area, and $\bar{v}_f$ is the average fluid velocity. If the channel is short relative to its width ($h_c/w_c \ll 1$), which is typically the case, and if we ignore the variation in velocity along the width of the channel (i.e. along the $z'$-axis), then the velocity profile reduces to,

$$v_f(y') = \frac{3\bar{v}_f}{2}\left[1 - \left(\frac{y'}{h_c}\right)^2\right]. \tag{38}$$

In order to include this expression in our analysis we rewrite it in terms of the coordinate $y$ of Fig. 2 in which $y' = y - (h + h_c + t_b)$ where $t_b$ is the thickness of the base of the channel (i.e., the distance from the top of the magnetic elements to the lower edge of the fluid). This gives,

$$v_f(y) = \frac{3\bar{v}_f}{2}\left[1 - \left(\frac{y - (h + h_c + t_b)}{h_c}\right)^2\right]. \tag{39}$$

Finally, we substitute Eq. (39) into Eq. (3) and obtain the fluidic force components

$$\mathbf{F}_{fx} = -6\pi\eta R_p \left[ v_x - \frac{3\bar{v}_f}{2}\left[1 - \left(\frac{y-(h+h_c+t_b)}{h_c}\right)^2\right]\right], \qquad (40)$$

and

$$\mathbf{F}_{fy} = -6\pi\eta R_p v_y. \qquad (41)$$

**E. Equations of Motion**

The equations of motion for a magnetic particle traveling through the microsystem can be written in component form by substituting Eqs. (33), (34), (40) and (41) into Eq. (1),

$$m\frac{dv_x}{dt} = F_{mx}(x,y) - 6\pi\eta R_p \left[v_x - \frac{3\bar{v}_f}{2}\left[1 - \left(\frac{y-(h+h_c+t_b)}{h_c}\right)^2\right]\right], \qquad (42)$$

$$m\frac{dv_y}{dt} = F_{my}(x,y) - 6\pi\eta R_p v_y, \qquad (43)$$

$$v_x(t) = \frac{dx}{dt}, \qquad v_y(t) = \frac{dy}{dt}. \qquad (44)$$

Equations (42) - (44) constitute a coupled system of first-order ordinary differential equations (ODEs) that are solved subject to initial conditions for $x(0)$, $y(0)$, $v_x(0)$, and $v_y(0)$. These equations can be solved numerically using various techniques such as the Runge-Kutta method.

## III. RESULTS

We demonstrate Eqs. (21), (22), and (31) - (34) via application to hypothetical but practical microsystem. Specifically, we study the behavior of magnetite ($Fe_3O_4$) particles in a microsystem with a fluid channel that is 200 μm high, 1 mm wide, and 10 mm long ($h_c$ = 100 μm, $w_c$ = 500 μm, and L = 10 mm). The fluid is nonmagnetic ($\chi_f = 0$), and has a viscosity and density equal to that of water, $\eta = 0.001$ N·s/m$^2$, and $\rho_f = 1000$ kg/m$^3$

· There is an array of 10 permalloy (78% Ni 22% Fe, $M_{es} = 8.6 \times 10^5$ A/m) elements embedded immediately beneath the microchannel.[17] Each element is 100 μm high, and 100 μm wide ($h$ = 50 μm, $w$ = 50 μm), and they are spaced 50 μm apart (edge to edge). Throughout this analysis we assume that the bias field is 0.25 T, which is provided by a single NdFeB magnet positioned beneath the microsystem.

The Fe$_3$O$_4$ particles have the following properties: $R_p$ = 250 nm, $\rho_p = 5000$ kg/m$^3$, and $M_{sp} = 4.78 \times 10^5$ A/m. We adopt a magnetization model for Fe$_3$O$_4$ described by Takayasu et al.[15], which is consistent with (13) when $\chi_p \gg 1$, i.e.,

$$f(H_a) = \begin{cases} 3 & H_a < M_{sp}/3 \\ M_{sp}/H_a & H_a \geq M_{sp}/3 \end{cases}, \qquad (45)$$

which implies that below saturation $\mathbf{H}_{demag} \approx \mathbf{H}_a$ ($\mathbf{H}_{in} \approx 0$), and $\mathbf{M}_p = 3\mathbf{H}_a$.

Throughout this analysis we use a coordinate system centered with respect to the first element (Fig. 4a). First, we compute the field and force along a horizontal line at two different heights y = h + 50 μm, h + 100 μm (50 μm and 100 μm above the elements). Plots of $B_x$, $B_y$, $F_x$ and $F_y$ are shown in Figs. 5-8, respectively. The field component $B_x$ and $B_y$ are for the magnetized elements only, and do not include the bias field. Notice that $B_x$ (Fig. 5) peaks near the edges of the element ($x = \pm w$) and alternate in sign from one edge to the other, whereas $B_y$ (Fig. 6) obtains its maximum value at the center of the element ($x = 0$). The horizontal force $F_x$ (Fig. 7) has a profile similar to $B_x$, but with an opposite polarity of sign. Thus, as a particle moves from left to right above an element it experiences a horizontal acceleration as it passes the leading edge, followed by a deceleration as it passes the trailing edge.

The profile of the vertical force component $F_y$ (Fig. 8) changes with distance above the element. Specifically, close to the element $F_y$ exhibits dual minima (attractive force) near the edges of the element.[19] However, farther away, a single minimum occurs at the center of the element (Fig. 8b). One of the most interesting aspects $F_y$ is that it alternates in polarity (direction) across an element. Specifically, it is positive (upward) just to the left of an element, negative (downward) directly above an element, and positive (upward) just to the right of an element. Therefore, as a particle moves across an element it first accelerates upward, then downward, and then upward again as it passes the element. This behavior is due to the nature of the force that the particle experiences, which, in turn, is due to the bias field and the collective field of the magnetized elements. In particular, the bias field tends to impose an upward vertical polarization in the particle throughout its range of motion, whereas the magnetized elements produce a field gradient that changes in both magnitude and sign pointwise within the microchannel. This gives rise to a spatially varying force on the particle that is attractive in some regions and repulsive in others. Thus, the particle exhibits oscillatory motion as it moves through the microsystem as shown below.

We now study the behavior of a particle as it moves through the microsystem. We assume that the particle enters the microchannel to the left of the first element at $x(0) = -10w$, and compute its trajectory as a function of its initial height above the magnetized elements: $\Delta y$ = 10 μm, 20 μm, …, 140 μm (i.e., initial heights of y(0) = 60 μm, 70 μm, …, 190 μm). The average fluid velocity is $\bar{v}_f = 10$ mm/s, and the particle enters the channel with this velocity, $v_p(0) = 10$ mm/s. As above, the bias field is 0.25 T.

The computed particle trajectories are shown in Fig. 9. It is easy to identify each trajectory with its entry height as this is the y-intercept for that plot. For this analysis we integrated Eqs. (42) and (43) using an implicit time-stepping algorithm, and it took only a few minutes to complete the simulation. We also used a fourth order Runge-Kutta time-integration scheme, and found no difference in our results. According to the analysis, particles that enter the microchannel 0-130 μm above the magnetized elements will be captured, but those entering 130 μm and above will pass through the system. The capture time (the time it takes for the particle to reach the bottom of the microchannel where it is held in place) is plotted as a function of entry height in Fig. 10. This plot shows that particles entering at heights 0-100 μm above the elements will be captured within 130 ms. Notice that the capture time is minimum for particles that enter the channel 50-60 μm above the elements. Particles that enter at lower heights ($\Delta y = 0$ ,.., 40 μm) experience a substantial vertical repulsive (upward) force prior to reaching the first element, which tends to extend their travel distance thereby increasing their capture time. Particles that enter at higher heights ($\Delta y = 70$ ,…, 140 μm) bypass the first element, and therefore have extended trajectories and longer capture times.

The capture efficiency of this system can be estimated as follows. If we assume that the particles are uniformly distributed height-wise as they enter the microchannel, then the percentage of particles captured will be equal to the maximum predicted capture entry height 130 μm, divided by the total height of the microchannel, which is 200 μm. Based on this analysis, approximately 65% of the particles that enter the microsystem will be captured. It should be noted that the dimensions and parameters used in this analysis were chosen to simply demonstrate the theory, and were not intended to optimize the capture efficiency. However, the capture efficiency for a proposed microsystem can be optimized prior to fabrication by performing a parametric analysis to determine the dimensions, spacing, and number of soft-magnetic elements that optimize capture for a given particle, microchannel, and flow rate. In principle, a capture efficiency of 100% can be achieved by increasing the size and number of soft-magnetic elements, or by reducing the height of the microchannel.

Finally, the microsystem described above has potential for a variety of applications especially in the field of microbiology where it can be used to transport, sort, and separate functionalized magnetic micro/nano-particles with surface-bound biomaterial. Our analysis indicates that the magnetic force generated by this system is on the order of 100 pN, which is more than sufficient to effect particle capture in applications such as bioseparation where the fluidic force is much lower. Furthermore, it might be possible to use this system to study fluid properties at the micro and nanoscale by sensing the oscillatory motion of the magnetic particles as they pass through it. This motion gives rise to a time varying magnetic field which could potentially be sensed using an integrated magnetometer. The magnitude of and period of the sensor signal could be used to determine the fluid properties such as viscosity, which influences the drag and hence the motion of the particle.

## IV. CONCLUSION

We have studied magnetophoretic particle transport and capture in a novel microsystem that consists of an array of integrated soft-magnetic elements embedded in a nonmagnetic substrate beneath a microfluidic channel. The elements are polarized using a bias field, and produce a force on magnetic micro/nanoparticles as they traverse the microchannel. We have derived the equations of particle motion using analytical expressions for the dominant magnetic and fluidic forces on the particles. The theory applies to dilute suspensions of particles for which interparticle effects are negligible. It takes into account the size and properties of the particles, the bias field, the dimensions of the microchannel, the fluid properties, and the flow velocity. We have used the theory to study particle transport and capture in a practical microsystem, and our analysis demonstrates the viability of using the microsystem for a variety of bioapplications such as bioseparation. Furthermore, the spatially periodic nature of the force within the microchannel could be used for various other applications including the characterization of the transport fluid at the micro or nanoscale.

**Figure Captions**

FIG. 1. Magnetophoretic microsystem: (a) microsystem with bias field structure, (b) cross-section of microsystem showing the microchannel, bias field, and magnetized elements.

FIG. 2. Geometry and reference frame for the microfluidic channel.

FIG. 3. Bias field analysis: (a) bias magnet and reference frame, (b) bias field distribution midway in the gap.

FIG. 4. Magnetized soft-magnetic elements: (a) 0'th element with reference frame, (b) array of elements with reference frame.

FIG. 5. $B_x$ above the magnetized elements: (a) $\Delta y = 50$ μm above the elements, (b) $\Delta y = 100$ μm above the elements, (c) cross-section of upper half of the element.

FIG. 6. $B_y$ above the magnetized elements: (a) $\Delta y = 50$ μm above the elements, (b) $\Delta y = 100$ μm above the elements, (c) cross-section of upper half of the element.

FIG. 7. $F_{mx}$ above the magnetized elements: (a) $\Delta y = 50$ μm above the elements, (b) $\Delta y = 100$ μm above the elements, (c) cross-section of upper half of the element.

FIG. 8. $F_{my}$ above the magnetized elements: (a) $\Delta y = 50$ μm above the elements, (b) $\Delta y = 100$ μm above the elements, (c) cross-section of upper half of the element.

FIG. 9. Particle trajectory vs. entry height above the magnetized elements.

FIG. 10. Particle capture time vs. entry height above the magnetized elements.

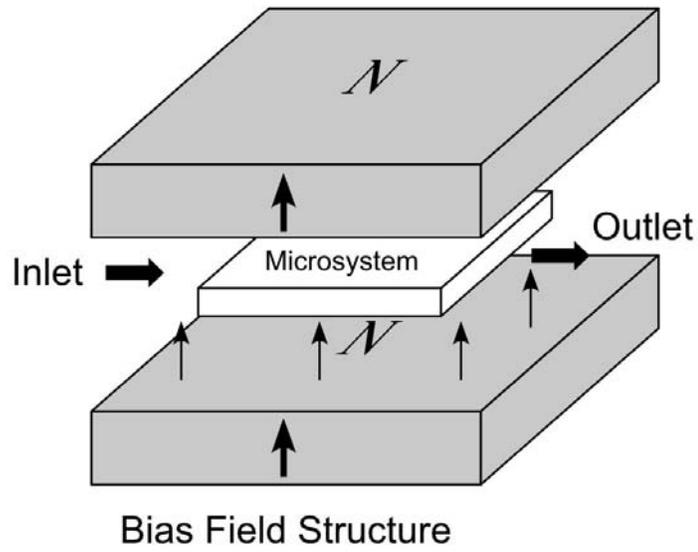

(a)

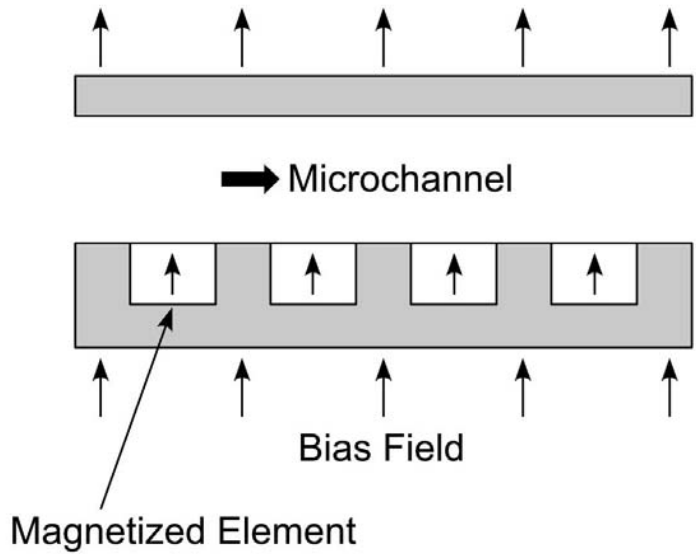

(b)

FIG. 1

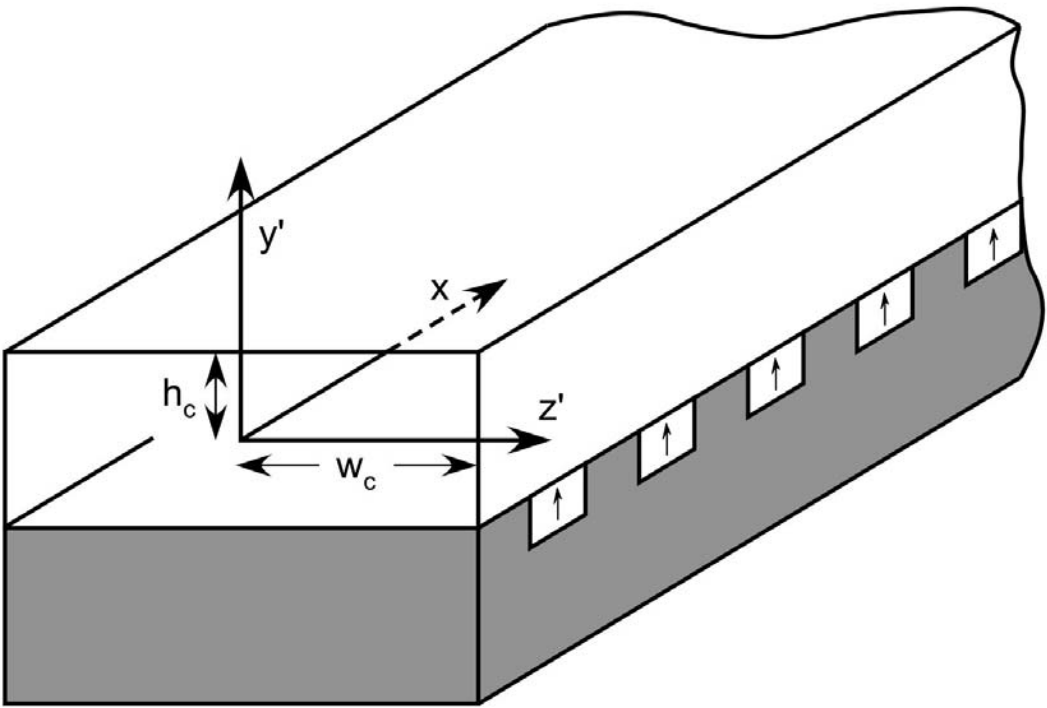

FIG. 2

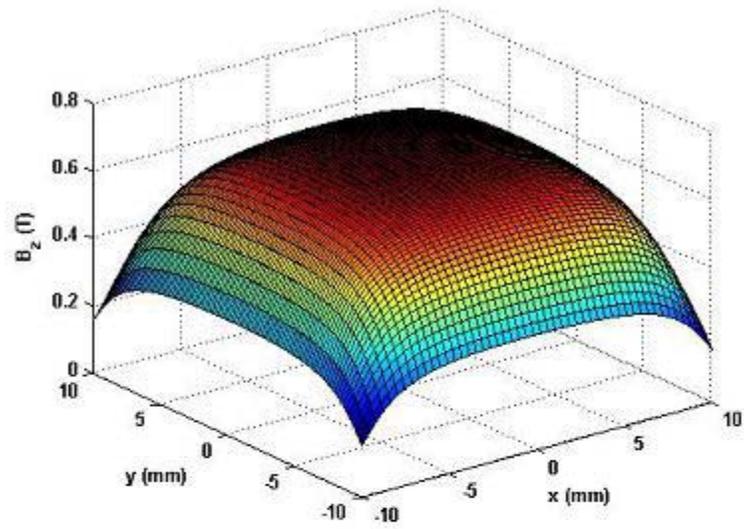

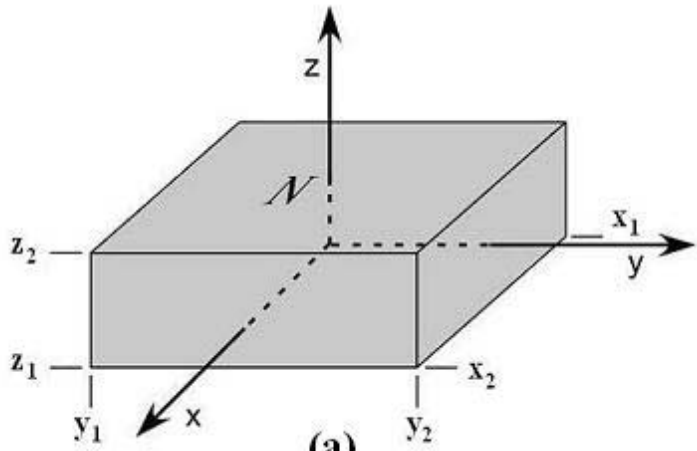

FIG. 3

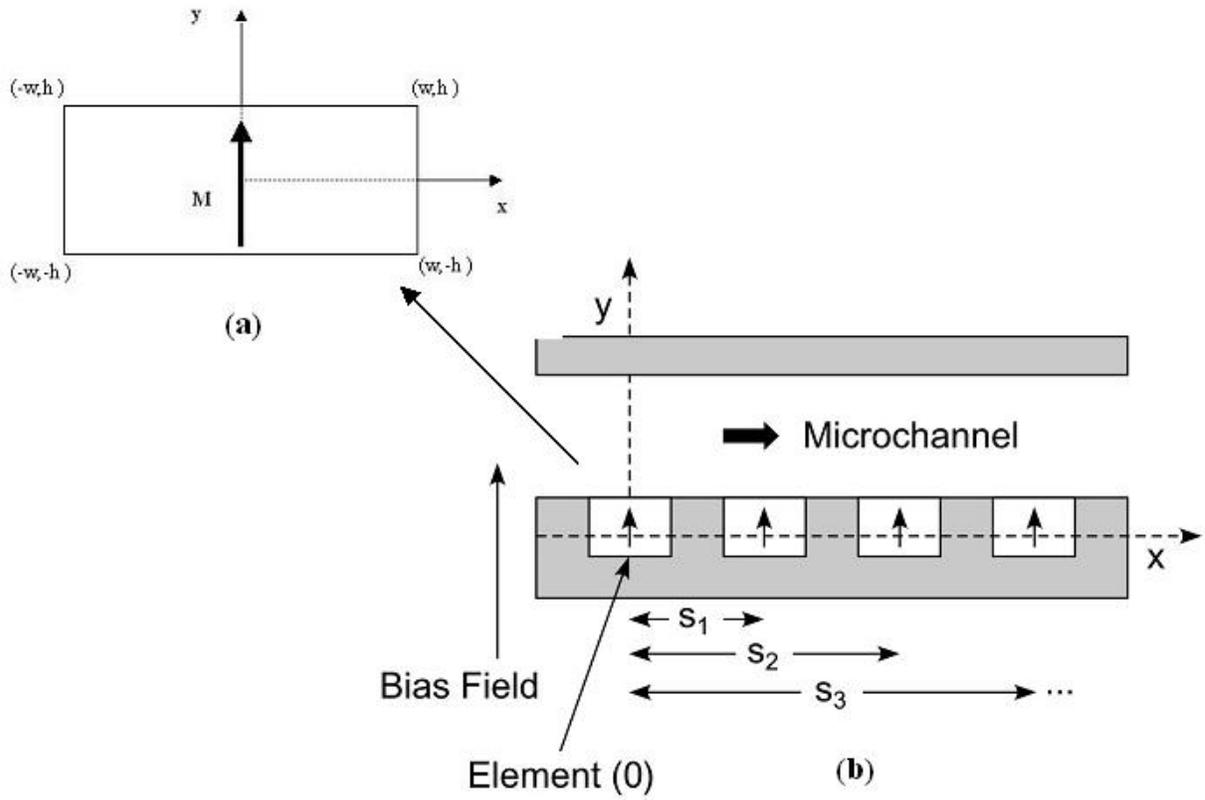

FIG. 4

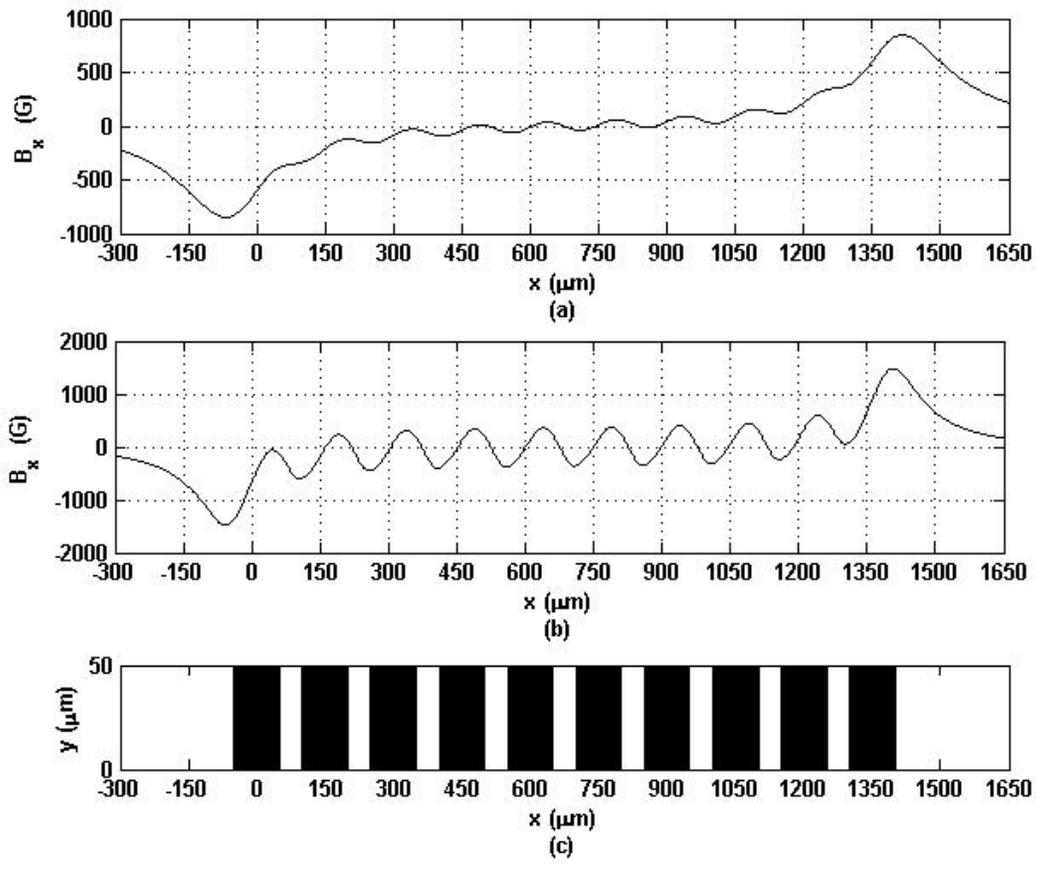

FIG. 5.

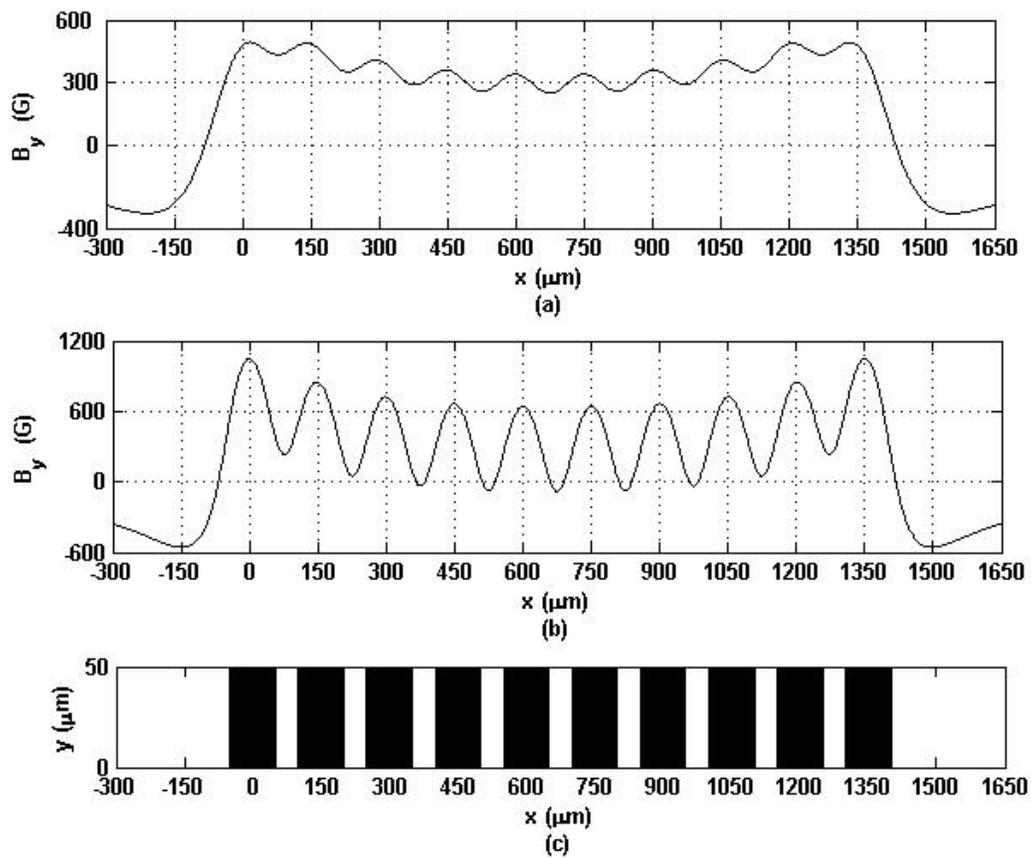

FIG. 6

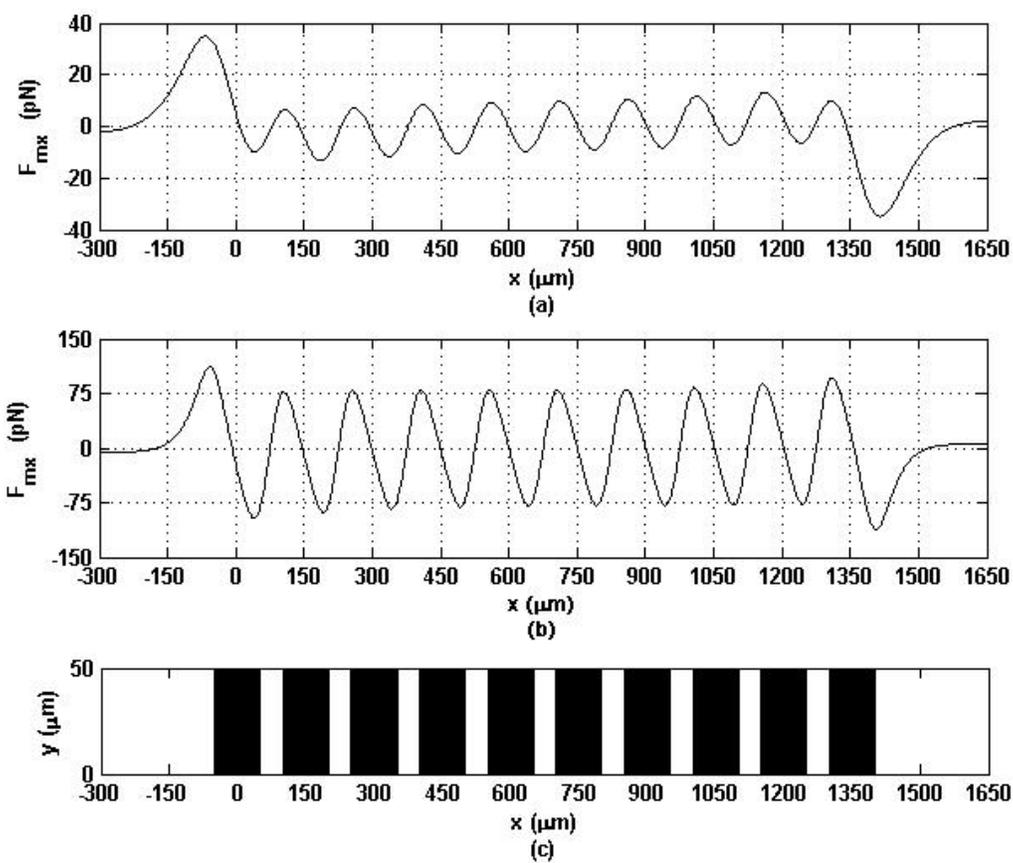

FIG. 7

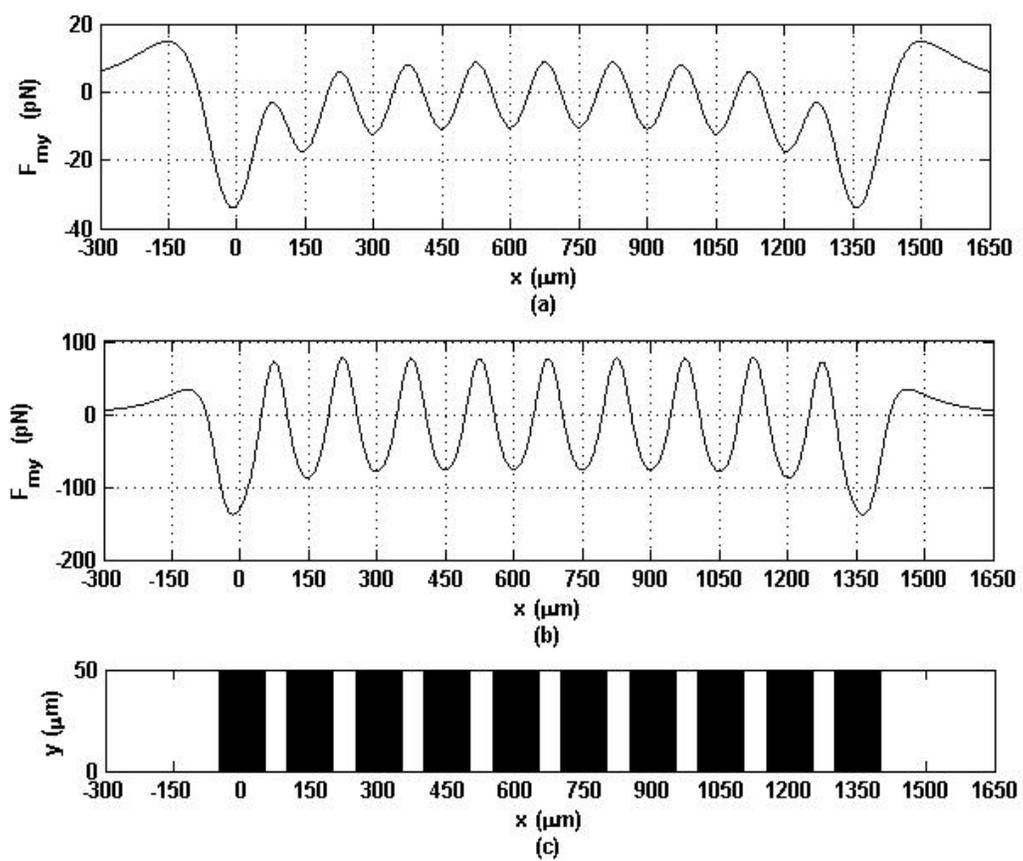

FIG. 8

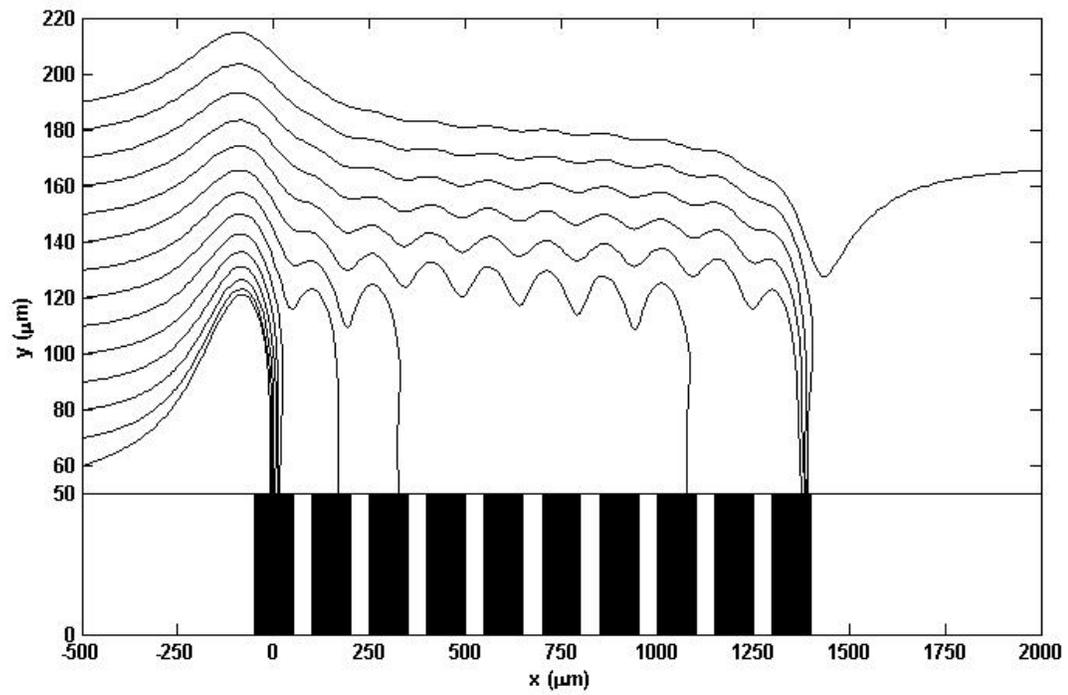

FIG. 9

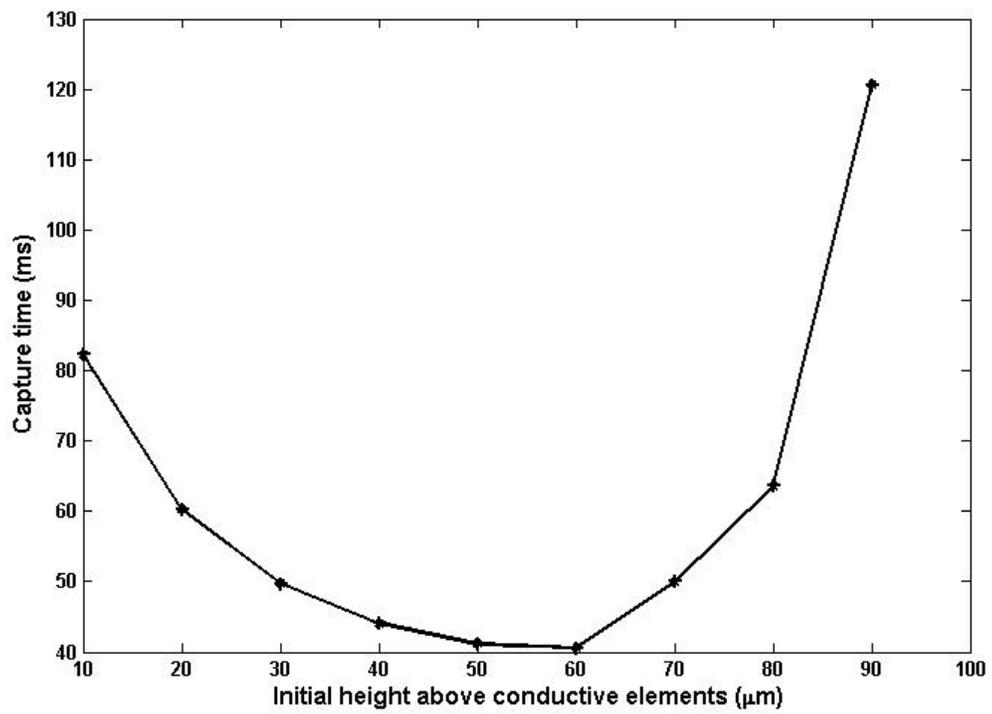

FIG. 10